\documentclass[usenatbib]{mnras}
\usepackage[T1]{fontenc}
\usepackage{aecompl}
\usepackage{amsmath,amsfonts,amssymb}
\usepackage[pdftex]{graphicx}
\usepackage{aas_macros}
\usepackage[usenames]{xcolor}
\usepackage{color}
\usepackage{textcomp,gensymb}
\usepackage{times}
\usepackage{subfig}
\usepackage{stfloats}
\usepackage[utf8]{inputenc}

\newcommand{\newc}[1]{{#1}}

\title[An Innie or an Outie?]{Are the spiral arms in the MWC 758 protoplanetary disc driven by a companion inside the cavity?}

\author[Calcino et al.]{
Josh Calcino,$^{1}$\thanks{E-mail: j.calcino@uq.edu.au}
Valentin Christiaens,$^{2}$
Daniel J. Price,$^{2}$
Christophe Pinte,$^{2}$
\newauthor{
Tamara M. Davis,$^{1}$
Nienke van der Marel,$^{3}$ and 
Nicol\'as Cuello$^{4}$}
\\
$^{1}$School of Mathematics and Physics, The University of Queensland, QLD 4072, Australia\\
$^{2}$School of Physics \& Astronomy, Monash University, VIC 3800, Australia\\
$^{3}$Department of Physics and Astronomy, University of Victoria, 3800 Finnerty Road,Victoria, BC, V8P 5C2, Canada \\
$^{4}$Univ. Grenoble Alpes, CNRS, IPAG, F-38000 Grenoble, France.
}

\date{Accepted XXX. Received YYY; in original form ZZZ}

\pubyear{2020}

\begin{document}
\label{firstpage}
\pagerange{\pageref{firstpage}--\pageref{lastpage}}
\maketitle

\begin{abstract}
Spiral arms in protoplanetary discs are thought to be linked to the presence of companions. We test the hypothesis that the double spiral arm morphology observed in the transition disc MWC~758 can be generated by an $\approx 10$ M$_{\rm Jup}$ companion on an eccentric orbit internal to the spiral arms. Previous studies on MWC~758 have assumed an external companion. We compare simulated observations from three dimensional hydrodynamics simulations of disc-companion interaction to scattered light, infrared and CO molecular line observations, taking into account observational biases. The inner companion hypothesis is found to explain the double spiral arms, as well as several additional features seen in MWC~758 --- the arc in the northwest, substructures inside the spiral arms, the cavity in CO isotopologues, and the twist in the kinematics. Testable predictions include detection of fainter spiral structure, detection of a point source south-southeast of the primary, and proper motion of the spiral arms.
\end{abstract}





\section{Introduction}
Where is the companion driving the spiral arms in the MWC 758 circumstellar disc? While spiral arms have long been expected from the theory of planet-disc interaction \citep{goldreich1979,goldreich1980}, recent high resolution images of circumstellar discs have shown a remarkable array of spiral arm morphologies \citep{garufi2017, dong2018c}. The disc around MWC 758 is one of the most spectacular --- with two prominent spiral arms seen in scattered light \citep{grady2013, benisty2015} and tentative evidence for a third spiral arm and 
a point-like source located roughly 20 AU from the central star \citep{reggiani2018}. 



\citet{dong2015b} first proposed that the spiral arms in MWC~758 are the result of a massive companion orbiting externally to the spiral arms (see also \citealt{fung2015,baruteau2019}). This outer companion alone does not explain the existence of the central cavity depleted in gas and large dust grains in MWC~758 \citep{boehler2018,dong2018}, and so cannot be the complete story. To explain the cavity and dust asymmetry, \cite{baruteau2019} proposed two companions, one inside the spiral arms, and one outside. Each companion carves a gap that triggers the Rossby Wave Instability. Their outer companion produces a vortex inside of its orbit, creating the dust asymmetry in the northwest seen in continuum observations \citep{boehler2018,dong2018}, while their inner companion produces a vortex outside of its orbit, creating the dust asymmetry seen in the southwest.

The comparison between the spiral arm morphology in MWC~758 with that produced by an external companion is compelling \citep{dong2015b}. The problem is that there has been no confirmed detection of such a companion. \cite{wagner2019} claimed detection of a 2-5 M$_\textrm{J}$ companion, but the projected separation of this object to the central star of $\sim$0.62 arcseconds placed it close to --- or even within --- the mm continuum emission. Such a massive body should be carving a deep, wide gap in gas and dust, but no such gap is seen in either \citet{boehler2018, dong2018}. However, the CO isotopologue observations reveal a deep gas cavity with an outer edge of 60 au, suggesting the presence of a planet in the inner part of the disc \citep{boehler2018}. The detection limits reported in \cite{wagner2019} are also in tension with the scenario presented by \cite{baruteau2019}, as their external companion of 5 M$_\textrm{J}$, located at 140 au, is within the claimed detection limits of \cite{wagner2019}, depending on the assumed start model.

In this paper we explore the hypothesis that the companion driving the spiral arms is located inside the gas and dust cavity. \cite{ren2018} claimed that this scenario is not possible in MWC~758 due to the measured rate of rotation of the spiral arms. However, the analysis by \cite{ren2018} is only valid for companions on circular orbits. Outward propagating spiral arms generated by planets also have low pitch angles, in conflict with the large pitch angles observed in MWC~758. But again, this is only true if the companion is on a circular orbit.

Low mass ($\lesssim $1~M$_\textrm{Jup}$) planets in protoplanetary discs are expected to be on nearly circular orbits since any initial eccentricity is damped by co-rotational and co-orbital eccentric Lindblad resonances \citep{goldreich1980,ward1986,artymowicz1993,papaloizou2000,tanaka2004}. This is not the case for massive planets that carve deep, wide gaps in the gas. A deep, wide gap allows the external Lindblad resonances to influence the dynamics of the gap opening body. Modest eccentricities on the order of \newc{$e \sim 0.2$ can develop \citep{papaloizou2001, ragusa2018}, depending on the mass of the companion and the disc properties. The mass threshold at which the eccentricity of the planet grows substantially depends also on the disc properties, and importantly, whether the planet is allowed to accrete material \citep{D'Angelo2006}.}

\newc{\citet{Muley2019} recently highlighted the impact of accretion on the evolution of massive planets' eccentricity}. 
Once the planetary companion has reached a mass ratio $q=M_{\mathrm{planet}}/M_{\mathrm{star}} \geq 0.003$, it experiences a sudden growth in eccentricity up to $e\sim 0.3$. \cite{Muley2019} were attempting to explain the tension between the observed position of PDS~70b with respect to the central star and the dust ring at 60 au \citep{long2018}. Although their hypothesis of an eccentric PDS~70b is now ruled out due to the detection of a second accreting body at a larger orbital distance than PDS~70b \citep{haffert2019,keppler2019,Wang2020}, this phenomenon may occur in other transition discs. How observationally inferred spiral arm morphology changes when the planet is on an eccentric orbit has not been widely explored.

\section{Methods}
We performed 3D simulations using the {\sc Phantom} smoothed particle hydrodynamics code \citep{phantom2018}. Our setup consists of a central protostar and a low mass companion in a coplanar orbit, both of which are modelled as sink particles with accretion radii of 2 au and 0.5 au, respectively \citep{bate1995}. The star and companion are free to accrete material, but the amount of material accreted is insignificant compared to the mass of either over the timescale of our calculations. The sink particles feel their mutual gravitational attraction, as well as the gravitational interaction of the gas. \newc{Hence, our companion is free to migrate in the disc and its eccentricity can evolve. Our simulations are run for approximately 250 orbits of the companion, however the spiral arm structure reaches a quasi-steady state over a few dozen orbits}.

\subsection{Simulation Initial Conditions}
The gas disc was set up in an annulus around the star and companion containing $2\times 10^{6}$ SPH particles with surface density $\Sigma(R)$ set as a power-law with $\Sigma (R) \propto R^{-p}$ for $R_\text{in} \leq R \leq R_\text{out}$, where we set $R_\text{in} = 47$ au, $R_\text{out} = 140$ au, and $p=1.8$. This is a rather steep density profile, but was motivated by the steep decrease in the density profile as derived from the intensity profiles of $^{13}$CO and C$^{18}$O seen in MWC 758 \citep{boehler2018}. Since we are assuming that our companion has already accreted most of its final mass and developed an eccentric orbit, the initial gas surface density starts with a cavity. \newc{The gas disc is set to be initially on a circular orbit. We set the Shakura-Sunyaev alpha viscosity to $\alpha_{\textrm{SS}} \approx 5\times 10^{-3}$ by employing a constant SPH artificial viscosity parameter $\alpha_{\rm AV} = 0.24$.} 

We set the total gas mass to $1.1 M_J$, consistent with the disc mass found by \cite{boehler2018}. As the disc evolves, the gas surface density quickly deviates away from this idealised initial setup. The thermal structure was assumed to be a power-law with $T(R) \propto R^{-q}$ where we set $q = 0.25$. The scale height of the disc was set to $H/R_{\text{ref}} = 0.05$ at $R_\text{ref} = 47$ au. The mass of the central sink particle was set to 1.5 $M_\odot$, in line with the mass of the central star in MWC 758 \citep{ancker1998}. The mass of the companion was set assuming a mass ratio $M_{\mathrm{planet}}/M_{\mathrm{star}} = 0.006$, giving a mass $M_{\mathrm{planet}} = 9.47\  M_{\mathrm{J}}$. We set the companion with a semi-major axis of $a = 33.5$ au and an eccentricity of $e = 0.4$. \newc{At the end of the 250 orbits, the eccentricity of the companion decreased to $e=0.388$, though most of this decrease occurs in the first 100 orbits. This is likely due to two reasons: the planet exchanging eccentricity with the disc, which is initiated on a circular orbit, and co-orbital Linblad resonances acting on the companion at apastron, when the companion collides with the cavity edge.}

\newc{Our choice of eccentricity is higher than that seen in \citet{Muley2019}. The growth of the eccentricity of the planet depends on a number of factors, particularly the accretion history. Further study of the eccentricity evolution of accreting bodies is required to validate our scenario. However we note that many of the morphological features we study in this paper can be achieved by a planet with a slightly lower eccentricity ($e\sim 0.3$), although we obtain a better match to the observed morphology of MWC~758 with a higher eccentricity.}


\subsection{Radiative Transfer}
 We made synthetic observations of our SPH simulation using the Monte Carlo radiative transfer code {\sc mcfost} \citep{pinte2006,pinte2009}. A Voronoi mesh was constructed around the particles from the SPH simulation, which was used an input model to {\sc mcfost}.

Since we did not compute the migration and evolution of dust grains, the dust population in our radiative transfer calculations is assumed to follow the gas (i.e. there is no settling) and have a power-law grain size distribution $dn/ds \propto s^{-3.5}$ for $0.03\mu$m $\leq s \leq s_{\textrm{RT max}} $. When computing our 1.04~$\mu$m polarised intensity images, we assume $s_{\textrm{RT max}} = 0.5\ \mu $m, and $ s_{\textrm{RT max}} = 3.5 \ \mu $m  when computing our 3.8~$\mu$m images. We found this necessary since the scattering properties of the disc change substantially when large grains are included in the surface layers \citep{Dullemond2004}. This is a reasonable assumption since any large dust grains will settle towards the mid-plane and be depleted from the surface of the disc seen in near-infrared observations \citep{Weidenschilling1980}. 





The gas mass used in the radiative transfer calculations was the same as in our SPH simulations. The dust mass was computed assuming a global gas to dust ratio $\epsilon = \rho_\text{g}/\rho_\text{d}$ and assuming the above power-law grain size distribution with $s_{\min}=0.03~\mu$m to $s_{\max}=1$ mm. Note that the gas to dust ratio for the grains with sizes $0.03~\mu$m $\leq s \leq s_{\textrm{RT max}}$ is much lower than the gas to dust ratio of all dust species in the disc. Therefore when we specify a certain gas to dust ratio in this paper, it includes all dust grain sizes, and not just the grains with sizes $0.03~\mu$m $\leq s \leq s_{\textrm{RT max}}$ which are used in the radiative transfer calculation. We make this distinction since it allows for easier comparison with estimates of the gas-to-dust ratio of discs in the literature.

\newc{The dust optical properties (absorption and scattering opacities, and scattering matrices) are computed using the Mie theory, assuming spherical and homogeneous grains, composed of astronomical silicate \citep{weingartner2001}.} The dust and gas are assumed to be in thermal equilibrium. We set the temperature of the central star to 8200 K \citep{ancker1998, boehler2018} and the luminosity was matched to the estimated value of 15.3 $L_\odot$ by changing the radius of the star and assuming it emits as a blackbody. \newc{We do not include the radiative effects from the companion, or the emission from circumplanetary disc}. When generating our $^{13}$CO images we assumed a constant $^{13}$CO-to-H$_2$ abundance ratio across the disc of $2\times10^{-6}$. The temperature profile of the disc is computed assuming the grain-size distribution explained above, with $s_{\textrm{max}_1} = 3.8\ \mu$m, and a gas-to-dust ratio of 10:1. We find that this high value of the gas-to-dust ratio is required in order to explain the scattering of near-IR photon inside $^{13}$CO cavity (see section \ref{sec:gsd}).

We used $10^8$ Monte Carlo photon packets to compute the temperature and specific intensities. Images were then produced by ray-tracing the computed source function. When making a direct comparison with MWC~758 we assumed an inclination of $i = 21^{\circ}$, a position angle PA$=62^{\circ}$ \citep{boehler2018}, and a source distance of 160 pc \citep{gaia2018}. When comparing line emission observations we simply convolved our images with a Gaussian function to match \newc{the} beam size of the observation.

\begin{figure*}
    \centering
    \includegraphics[width=0.8\linewidth]{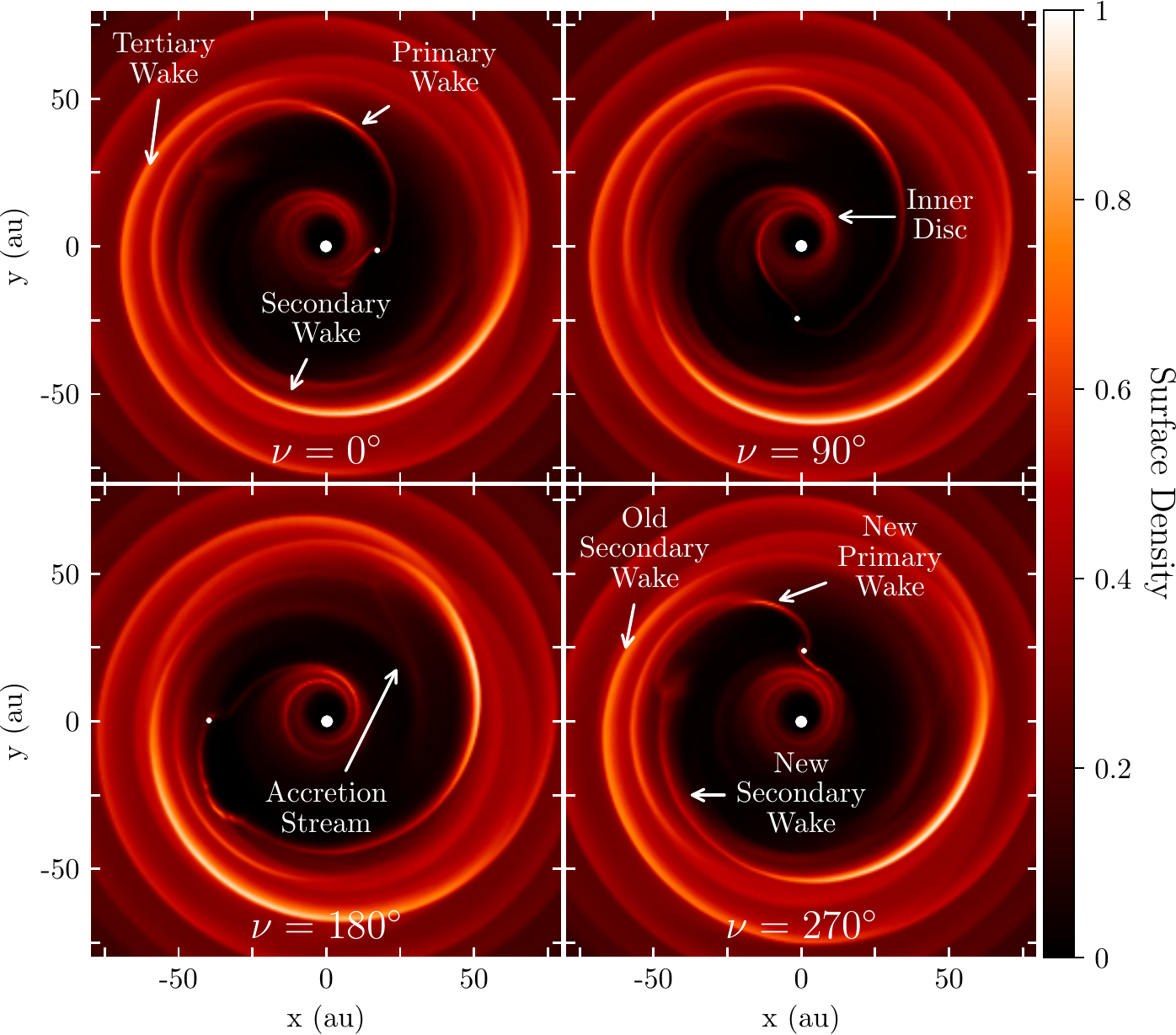}
        \caption{The surface density of the gas over a single orbit of the companion. A true anomaly $\nu=0^{\circ}$ corresponds to the position of the planet at periastron, while $\nu = 180^{\circ}$ is at apastron. The star and planet are marked with the large and small white circles, respectively. As the companion progresses through its orbit, the pitch angle of the primary and secondary outer wakes changes. At certain positions, the pitch angle of the primary outer wake can become quite high (e.g. $\nu = 0^{\circ}$).}
    \label{fig:sd_plot}
\end{figure*}

\begin{figure*}
    \centering
    \includegraphics[width=\textwidth]{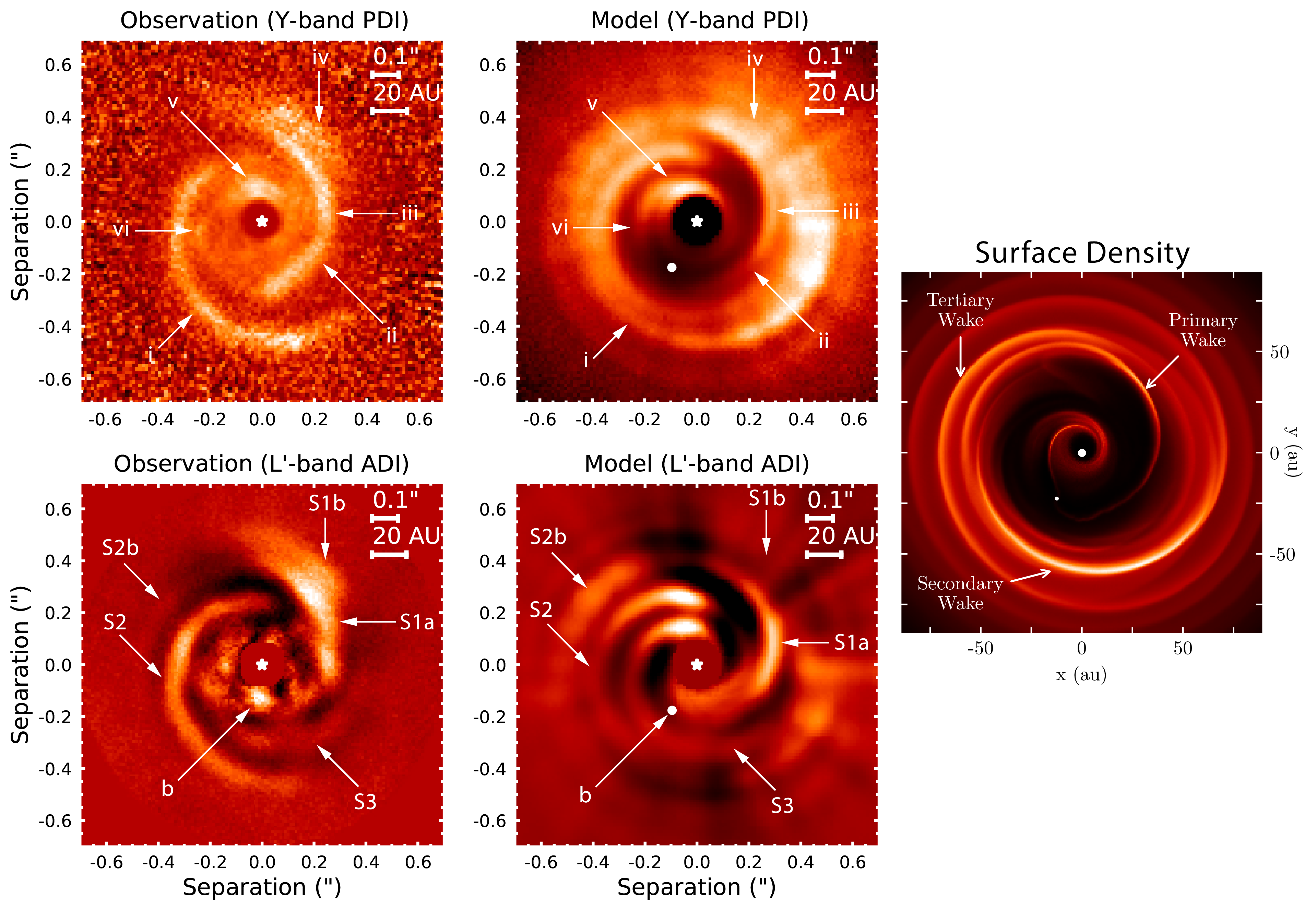}
    \caption{The comparison between the Y-band polarised intensity images obtained by \protect\cite{benisty2015} compared with the Y-band polarised intensity of a timestep of our simulation (top row). We have labelled the major features noted by \protect\citet{benisty2015} and the analogous structures in our model. We are able to reproduce the double spiral structure, as well as the arc in the northwest. The bottom row compares the L'-band images of MWC 758 from \protect\cite{reggiani2018}, with our L'-band model in the bottom right panel. \protect\cite{reggiani2018} claimed the detection of a third spiral (S3), which we reproduce in our model. Our companion is also in a similar location to the companion candidate (b), and is marked with a white circle. \newc{Note that we do not necessarily expect to see emission from the companion in the PDI images (top right panel), but still mark its location for convenience.} We also label S2b, an additional faint spiral structure outside of S2. \newc{The sky projected surface density of the simulation at the timestep when we match the observations is shown in the far right panel.}}
    \label{fig:qphi}
\end{figure*}

\begin{figure*}
    \centering
    \includegraphics[width=0.8\textwidth]{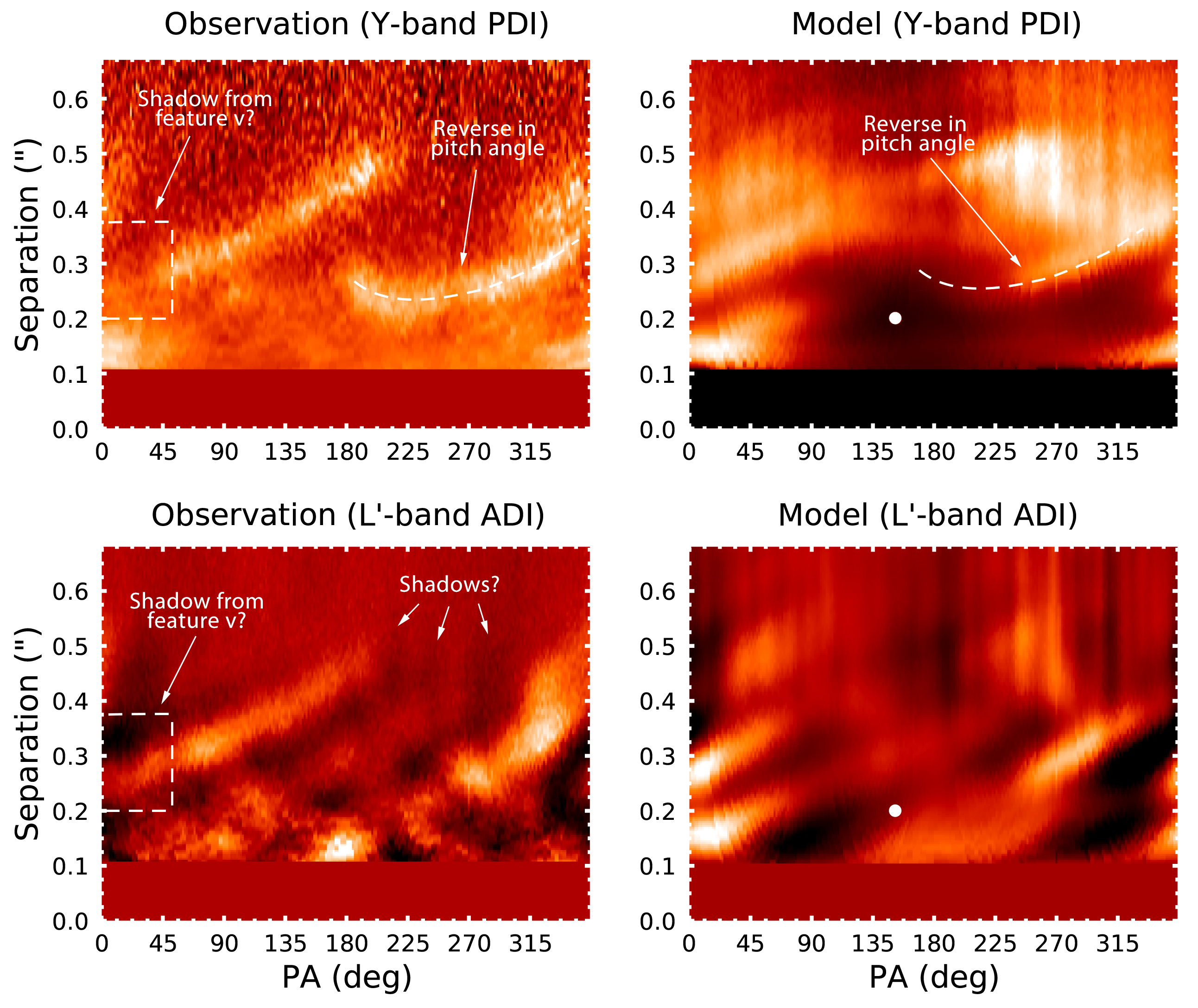}
    \caption{As in Figure \ref{fig:qphi}, except the images are plotted in polar coordinates. In the $Y$-band PDI images of \protect\cite[][top left panel]{benisty2015}, the spiral arm S2 (between $\sim 180^{\circ}$ to $\sim 360^{\circ}$) appears to suddenly reverse in pitch angle. We cover the spiral with a dashed line to better illustrate this feature, but note that this dashed line is just a qualitative match. Rather than continuously increasing in separation, S2 first begins to decrease in separation, before finally increasing again. This is qualitatively match in our $Y$=band PDI model (top right panel), however due to shadowing effects the tip of the spiral is fainter than in the PDI images of \protect\cite{benisty2015}. Note that the line plotted over S2 in our model is the same line is plotted over S2 in the observations. We also note that there appears to be shadowing from feature v (see Figure~\ref{fig:qphi}) onto the outer spirals.}
    \label{fig:rphi}
\end{figure*}

\subsection{Post-processing of Infrared Images}

Infrared observations at high contrast and high angular resolution involve the use of observing strategies and post-processing algorithms that can affect the flux and geometry of faint circumstellar signals (disc or planets).
The most sensitive observations of the vicinity of MWC~758 to date were obtained using the polarimetric differential imaging \citep[PDI;][]{Kuhn2001} and the angular differential imaging \citep[ADI;][]{Marois2006} techniques, applied to coronagraphic VLT/SPHERE and Keck/NIRC2 infrared observations \citep[][respectively]{benisty2015,reggiani2018}.
For a meaningful comparison to these observations, we have mimicked the following effects in our infrared synthetic observations:
\begin{enumerate}
    \item convolution with the observed point spread function; 
    \item pixel sampling; 
    \item transmission of the coronagraph;
    \item \newc{shot noise};
    \item for the PDI synthetic images: conversion to the azimuthal component of the polarisation 
    and $r^2$-scaling of each pixel intensity; 
    \item for the ADI synthetic images: image post-processing by an algorithm based on principal component analysis \citep[PCA;][]{Amara2012,Soummer2012}. 
\end{enumerate}

To implement these effects, we used routines of the Vortex Image Processing pipeline\footnote{Available at: \url{https://github.com/vortex-exoplanet/VIP}.} \citep[VIP;][]{GomezGonzalez2017}, an open-source compilation of high-contrast imaging routines written in Python.
For either the PDI or the ADI synthetic observations, we first convolved the images with the point spread function of VLT/SPHERE and Keck/NIRC2: \newc{59mas and 79mas FWHM in the $Y$-band \citep{benisty2015} and $L'$-band \citep{reggiani2018} observations, respectively}. We then resampled the synthetic images to the same plate scale as the observed images \newc{(12.5mas/px and 9.9mas/px for the SPHERE and NIRC2 images) using a fourth order lanczos interpolation}.
We subsequently applied the effect of each coronagraph: an apodized Lyot coronagraph for the VLT/SPHERE observations \citep{Guerri2011}, and a Vortex AGPM coronagraph for the ADI observations \citep{Mawet2005,Delacroix2013}.
The effect of the apodized Lyot coronagraph used in the observations of \citet{benisty2015} is reproduced with an opaque mask of 185 mas diameter in the center of the image and the off-axis transmission curve measured in \citet{Guerri2011}.
For the AGPM coronagraph of NIRC2, we considered a stellar attenuation following a negative gaussian with a peak rejection factor of 500 and FWHM set to 0.9 times the size of a resolution element ($\sim$80 mas for Keck/NIRC2 in $L'$ band), and the off-axis transmission curve reported in \citet{Serabyn2017}.
\newc{We then considered the effect of shot noise, by injecting random Poisson noise in our simulated images after scaling pixel intensities to a similar level as in the observed images.}

For the PDI observations, we computed the azimuthal component of the polarisation (expected to contain most disc signal) in the so-called $Q_{\phi}$ map, as in \citet{benisty2015}. $Q_{\phi}$ is computed from the Q and U linear polarisation components of the image provided by {\sc MCFOST}:
\begin{equation}
    Q_{\phi} = Q \cos(2\phi) + U \sin(2 \phi),
\end{equation}
where $\phi$ is the polar angle measured from the positive $x$ axis. Each pixel intensity is finally scaled by a factor $r^2$, where $r$ is the radial separation to the star, to account for the dilution of stellar flux and enhance details at large separation in the disc.

For the ADI observations, we considered the $L'$ total intensity {\sc MCFOST} image and created an ADI sequence of 80 frames spanning the same field rotation range ([$-128\deg$,$103\deg$]) as reported for the \citet{reggiani2018} observations, by rotating the image with the corresponding values. We then post-processed the cube with the same PCA-ADI algorithm and the same number of principal components as used in \citet{reggiani2018} to model and subtract the signal from the star.
While our procedure does not deal with speckle noise, our goal is to reproduce the geometric and flux biases induced by ADI on extended disc signals \citep[e.g.][]{Milli2012,Christiaens2019}.

\section{Results}

\subsection{Gas Surface Density}\label{sec:gsd}

 Figure~\ref{fig:sd_plot} shows the resulting surface density of the gas in our simulation over a single orbit of the companion shown after 60 orbits. The positions of the star and planet are marked with white circles matching the size of their accretion radii. 


Low mass planetary companions are expected to produce a single inner and outer wake in the gas surface density \citep{goldreich1979, goldreich1980, ogilvie2002}. This changes when the planetary mass increases to the thermal mass (the mass at which the Hill radius of the planet is comparable to the scale height of the disc). In this case, a primary and secondary inner/outer wake are produced \citep{zhu2015, dong2015b}. The secondary outer wake generated has a larger radial width and a lower amplitude than the primary arm \citep{juhasz2015, dong2015b}, however the amplitude increases with the mass of the companion \citep{juhasz2015}. When eccentricity is negligible, the secondary wake is shifted 180 degrees from the primary wake.

When $\nu = 0^{\circ}$, we can see one spiral arm directly connected to the companion, which we identify as the primary outer wake at this particular true anomaly. A secondary outer wake is also apparent, but has a narrower radial width and higher amplitude compared to the circular case \citep[e.g. see][]{zhu2015,dong2015b}. 

As the companion moves through its orbit, the pitch angle of the primary outer wake changes. This is in contrast with the case for a planet on a circular orbit, where the outer wake pitch angle remains constant during the orbit. The primary wake remains attached to the companion throughout the orbit until $\nu \sim 270^{\circ}$, where it appears that the primary and secondary outer wake exchange roles. So we see that what appears to be the secondary wake when $\nu = 0^{\circ}$ was the primary outer wake of the companion during its previous orbit. This explains why the secondary outer wake in $\nu = 0^{\circ}$ has a narrower azimuthal width and higher amplitude than expected. Therefore as the companion moves through its orbit, new spiral arms are periodically created. Part of the original secondary arm propagates outward, and becomes the `tertiary wake' labelled in Figure \ref{fig:sd_plot}.

Accretion streams, which do not appear to be connected to the companion, are also apparent inside the gas cavity, as in HD~142527 \citep{price2018}. The leading edge of the secondary arm appears to begin falling inside the cavity when $\nu \sim 0^\circ - 180^\circ$. This leading edge is then disconnected from the rest of the spiral when the companion begins pulling on the secondary when $\nu = 0^\circ$. 

The inner gas disc around the primary star is also worth discussing. This is the gas that is inside the orbit of the companion, circling the primary star. In our simulation this gas is on a slightly eccentric orbit, where its apastron is roughly in the northern direction (shifted roughly 90$^\circ$ degrees from the apastron of the companion). There is also a cavity inside this gas, close to the primary sink particle. This inner cavity is a result of the boundary condition for our sink particle; SPH particles that cross the accretion radius become accreted and their mass and angular momentum is added to the sink particle. If the accretion radius of the sink particle is reduced, then SPH particles can orbit the sink at a closer radius, but this increases the run-time for the simulation. Therefore, we expect that our inner disc would extend much closer to the primary star than is reflected in our simulations. \newc{Unfortunately we are not able to follow the evolution of this inner disc for a very long due to the accretion. Thus we cannot comment on whether this inner disc will freely precess, or if it is restricted by the orbit of the companion.}

\begin{figure*}
    \centering
    \includegraphics[width=0.8\textwidth]{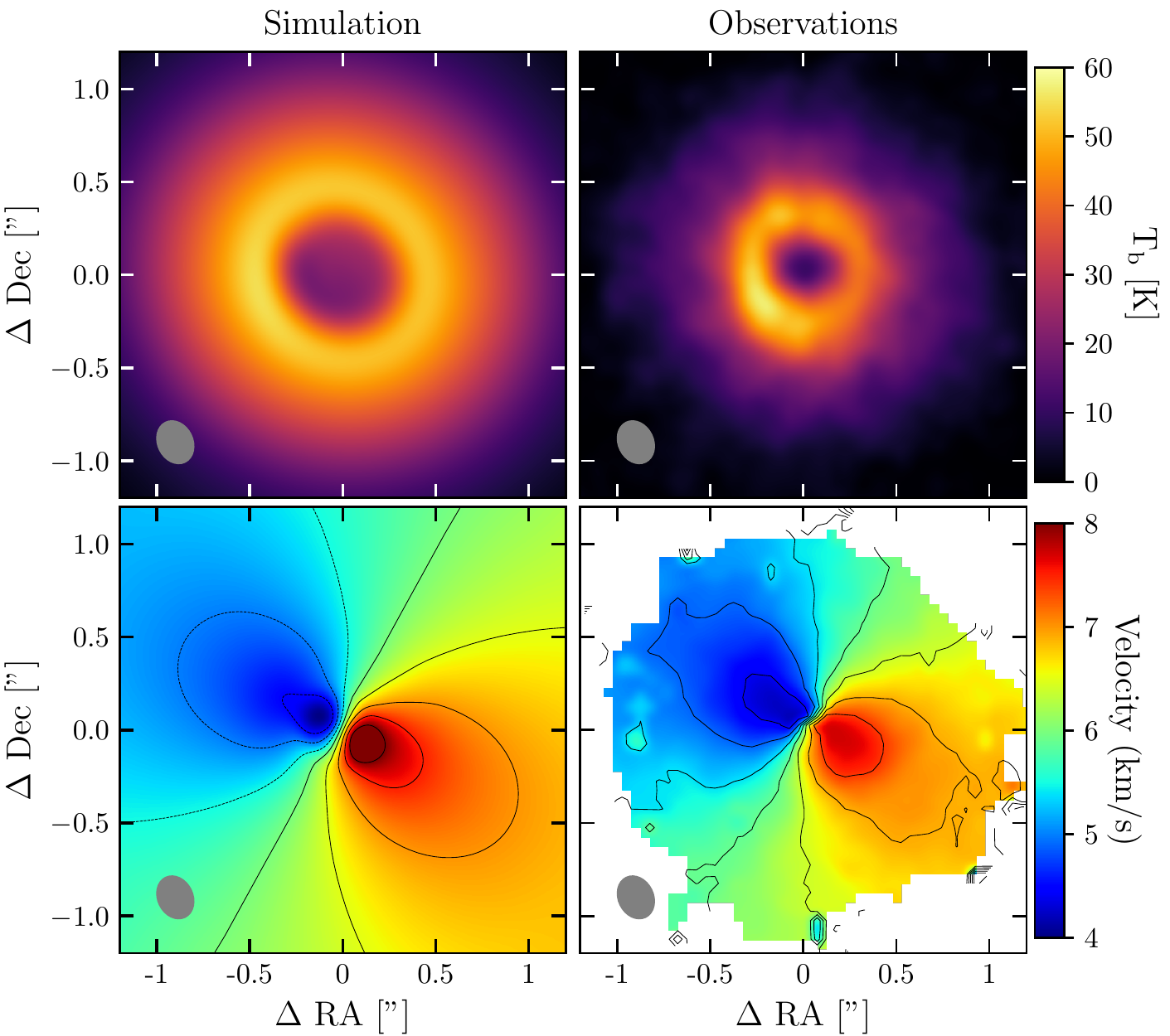}
    \caption{A comparison between the $^{13}$CO (3-2) integrated emission (top row) and velocity profile (bottom row) of our simulation with observations of MWC~758 presented in \protect\cite{boehler2018}. We recover the central cavity in $^{13}$CO with our eccentric companion. The high brightness temperature outside of the cavity indicates that the temperature, density, and/or $^{13}$CO abundance in this region is lower in MWC~758 than in our simulation.}
    \label{fig:co}
\end{figure*}

\subsection{Comparison with NIR Observations of MWC 758}\label{sec:com_nir}

The scattered light images of MWC 758 presented in \cite{benisty2015} show two grand spiral arms with several peculiar extra features. These are labelled in the top-left panel of Figure~\ref{fig:qphi}:
\begin{enumerate}
    \item a spiral in the southeast (\newc{which we} refer to as spiral S2),
    \item an arc in the southwest,
    \item a spiral in the northwest (referred to as spiral S1a),
    \item an arc located radially outward of S1a (referred to as spiral S1b),
    \item northern substructure inside the spirals,
    \item eastern substructure inside S1.
\end{enumerate}
We compare the $r^2$-scaled polarised intensity image of MWC~758 from \cite{benisty2015} to the polarised intensity image \newc{at one particular time of our simulation after 60 orbits of the companion} (top-right panel of Figure \ref{fig:qphi}), and use the same labels for corresponding features found in our simulation. \newc{We use a timestep relatively early in the evolution of our simulation since after prolonged evolution the inner disc becomes depleted of SPH particles, which is mainly due to the resolution of the simulation and size of the central sink particle.} 

\begin{figure*}
    \centering
    \includegraphics[width=0.8\textwidth]{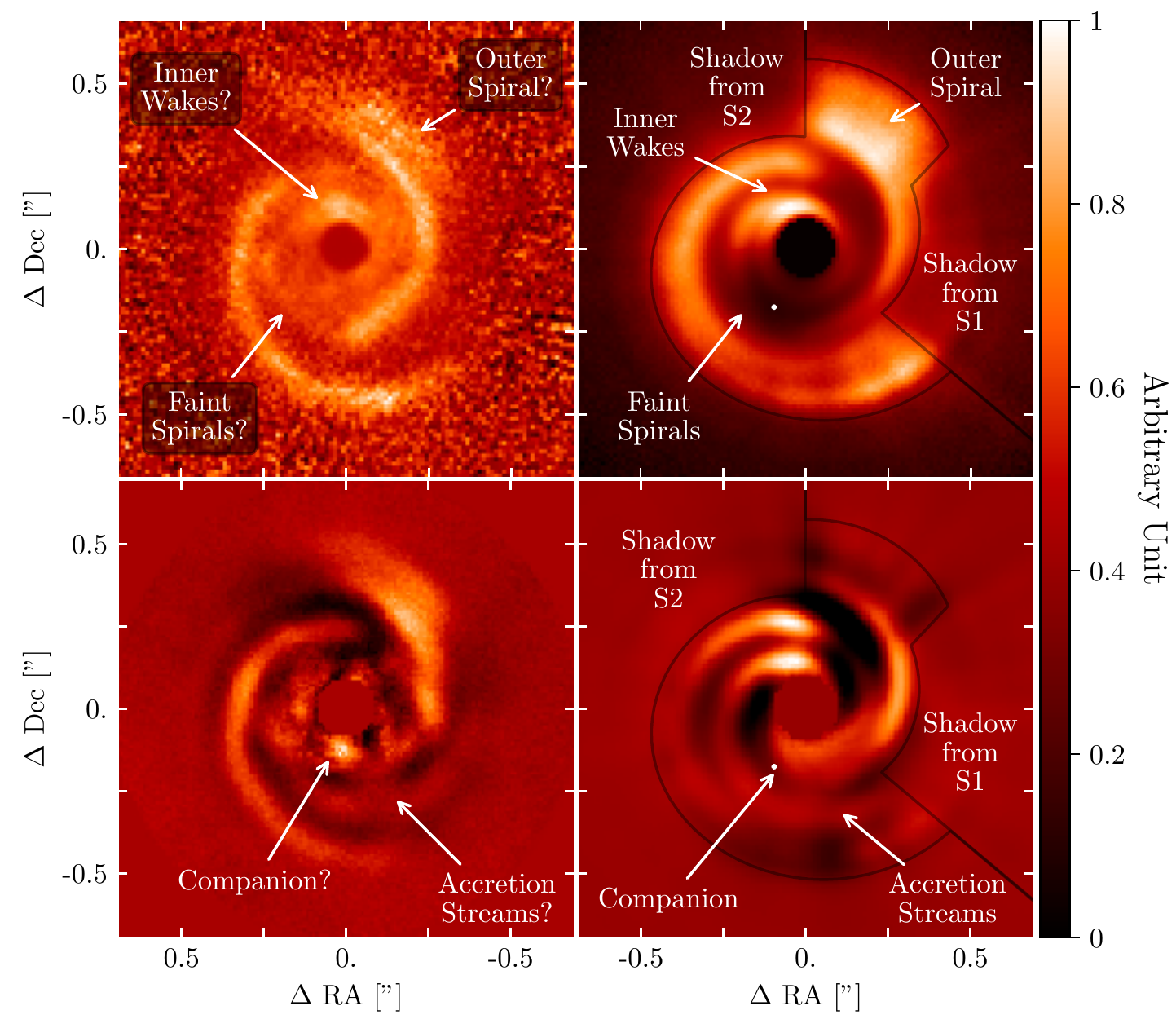}
    \caption{A comparison of the $Y$-band PDI observations from \protect\cite{benisty2015} (top left) and the $Y$-band PDI image of our simulation (top right), along with the $L'$-band ADI observations from \protect\cite{reggiani2018} (bottom left) are also shown and our $L'$-band ADI image. We propose that the outer regions of the disc are shadowed by S1 and S2, and that there is additional spiral structure inside the cavity. The spiral S3 arises from accretion streams into the cavity, and is the leading edge of the tertiary spiral shown in Figure \ref{fig:sd_plot} and the right panel of Figure \ref{fig:qphi}. The northwestern arc (S1b) is also part of this tertiary spiral (labelled ``Outer Spiral'' in this Figure). The companion candidate is in a close position to the companion in our simulations, however we remain cautious about its true nature.}
    \label{fig:dens_vs_scat}
\end{figure*}

We qualitatively reproduce all of the major features in the polarised intensity image of \cite{benisty2015}, and will detail the origin of each. \newc{The origin of each feature can be discerned from the surface density plot in the far right panel, which has been orientated to be in the sky projected plane}. The spirals S1a and S2 (features iii and i) are the primary and secondary outer wakes of the companion, respectively. Feature (ii) is then an extension of the S1a spiral. Feature (iv) is part of the tertiary spiral seen in the top left panel of Figure \ref{fig:sd_plot}, the majority of which is being shadowed by S1 and S2 (see Figure \ref{fig:dens_vs_scat}). Feature (v) in our simulation is the inner wakes of the companion, and the edge of the eccentric disc around the primary star, as discussed in Section~\ref{sec:gsd}.  Since our companion is substantially larger than the thermal mass, we expect that it to have multiple inner wakes. Feature (vi) may be part of the spiral structure leading the companion, or may be spiral structure colliding with the inner wall of the gas cavity. This effect can be seen when $\nu = 180^{\circ}$ in Figure \ref{fig:sd_plot}, where the primary outer wake of the companion collides with the cavity edge creating an enhancement in surface density. The enhancement in density increases the height of the scattering surface at this location in the disc, allowing it to intercept and scatter more photons.

The bottom row of Figure \ref{fig:qphi} shows our comparison to the L'-band ADI image of MWC 758 obtained by \cite{reggiani2018}. The two most interesting features found in this work are the additional spiral structure (S3) and the companion candidate (b). In our model, S3 arises due to accretion streams moving inside the cavity leading the tertiary spiral, which is also responsible for the arc in the northwest (labelled iv and S1b). Although tentative, S3 may be connecting to S1b in MWC 758, which is also occurring in our model.

We also label an additional faint spiral structure, S2b, which is not described in the L'-band ADI images of \cite{reggiani2018}, but appears in the total intensity images from \cite{grady2013}. In the L'-band images, S2b appears to be a continuation of S1a, but with reduced brightness due to shadowing from S2. We also see this in our L'-band ADI model. This spiral is the tertiary wake labelled in Figure \ref{fig:sd_plot}. 

\newc{We plot the observations and our model in polar coordinates in Figure \ref{fig:rphi} to better study changes in pitch angle of the spirals. A rather peculiar feature of S1 in the \cite{benisty2015} $Y$-band PDI images (see top left panel) is that it first decreases in radial separation with respect to the central star, before increasing for larger values of PA. We have plotted a dashed line over the top of S1 to illustrate our point, however this line is purely qualitative, and is not a fit to the spiral. 
We also obtain this feature, and include the same dashed line plotted on top of S1 in our model (top right panel). The inner tip of S1 in our model is shadowed by the inner disc, so does not appear as bright as in the observations. However we do also see this in the surface density plot in the far right panel of Figure \ref{fig:qphi}.}

\newc{Also noticeable in Figure \ref{fig:rphi} is the appearance of shadows in the $L'$-band ADI images (bottom left panel) from \cite{reggiani2018}. Shadowing effects are much clearer to see in polar coordinates since they appear as vertical striations. Shadowing is also apparent in both our $Y$-band and $L'$-band images. The shadows in our model are originating from optically thick portions of the spiral arms, and not due to inclination effects. This appears to be the case also in the observations, with the possible exception of the shadowing correlated with feature v, which could plausibly be due to a somewhat inclined inner disc.}\newc{What is also particularly interesting is that the bright inner material of the $Y$-band images (feature v) is co-located in position angle with shadowing at larger separations in the $L'$-band images. Closer inspection of S1 (between $\sim 0^{\circ}$ and $\sim 30^{\circ}$) in the $Y$ band shows that it may be shadowed behind feature v. This shadowing is evident on and behind S1 in the $L'$-band images. 
}



\subsection{CO Emission}\label{sec:co_em}

Figure~\ref{fig:co} shows the $^{13}$CO (3-2) moment~0 and moment~1 maps produced from the velocity field of our simulation after 200 orbits of the companion (left) compared to the observations in MWC~758 (right). The emission from the outer regions of the disc is higher than observed in MWC~758, and our cavity size is larger. This implies that either the temperature, density, or CO abundance (or a combination of these) in this region is higher in our simulation than in MWC~758. The cavity size in the C$^{18}$O observations from \cite{boehler2018} is larger than the $^{13}$CO. The same is also true for our simulated C$^{18}$O emission. This indicates the $^{13}$CO emission is more optically thick and therefore sensitive to the temperature profile. In Section \ref{sec:ilum} we discuss how illumination effects will change the temperature profile of the disc as the spiral arms are heated and shadow more distal regions of the disc, which may then make the inner edge of the gas cavity appear brighter in $^{13}$CO emission. We cannot rule out the possibility that a companion with a smaller semi-major axis is required to explain the smaller size of the $^{13}$CO and C$^{18}$O cavity. 

The moment-1 (velocity) map of MWC~758, shown in the bottom right panel of Figure \ref{fig:co}, shows hints of a departure from Keplerian rotation \newc{in the very central regions}. \cite{boehler2018} suggest that this deviation is due to either radial flows or a misaligned inner disc. The degree of misalignment of the 3 au inner dust disc was recently quantified by \cite{francis2020} to be roughly 30$^\circ$ with respect to the outer disc. The CO emission responsible for the inner twist in the velocity map of MWC~758 must be much larger than the 3 au inner dust disc, however this could be consistent with the radial drift of dust down to a 3 au scale compared with a larger inclined inner gas disc. \newc{Such an inclined inner disc would presumably cause a sharp double shadow feature on the outer disc, as in the case of HD 142527 \citep{marino2015} and HD 100453 \citep{benisty2017}. This double shadow feature is not seen in MWC~758 \citep{grady2013, benisty2015}.}

The velocity map of our simulation exhibits \newc{perturbations} inside the $^{13}$CO cavity\newc{, shown in the bottom left panel of Figure \ref{fig:co}. These perturbations are most obvious slightly south of the centre of our velocity map. Comparing with the velocity map of MWC~758 in the bottom left panel, we can see similar perturbations, albeit not in the same location as our perturbations. Since our companion is co-planar with respect to the disc, these perturbations are not originating due to warping of the disc, but rather from radial flows induced by the eccentric companion. Other works have shown the importance of the kinematics in inferring the presence of planetary \citep[e.g.][]{pinte2018,pinte2019} and stellar companions \citep[e.g.][]{calcino2019, poblete2020} in the disc.} Determining the origin of the twist in MWC~758 will help refine our model, as whether it originates due to an inclined inner disc, or radial flows, can imply different consequences for our companion (e.g. a somewhat inclined or more eccentric orbit).

\newc{We use a timestep of our simulation after 200 orbits of evolution to produce our moment~0 and moment~1 images since at this stage the central regions of the disc are more cleared than in earlier times (as in Figure \ref{fig:sd_plot}). Our CO moment~0 at 60 orbits of evolution shows $^{13}$CO emission inside the cavity owing to the large amount of gas around the central star. This disc viscously spreads and accretes onto the central sink as the simulation progresses, resulting in a lower mass central disc. Note that in SPH the effective viscosity is limited by the resolution of the simulation. Thus a poorly resolved disc is more viscous than higher resolution disc with the same SPH artificial viscosity parameter. A lower mass inner disc has a lower scattering surface than a more massive inner disc, which when producing synthetic scattered light images results in less scattering off this inner disc.}

\newc{This may be a challenge for our proposed scenario, but it also raises a rather peculiar feature of MWC~758. Despite the cavity appearing relatively depleted, there is still significant scattering inside the cavity, particularly with the appearance of feature v in the $Y$-band PDI images from \cite{benisty2015}, shown in the top left panel of Figure \ref{fig:qphi}.  One plausible way to overcome this challenge in our model is if our inner disc were lower mass but with a higher temperature in the SPH calculation, and hence have a larger scale height. This inner disc would then intercept more stellar radiation, but may not appear in $^{13}$CO observations. Obtaining higher spatial resolution observations of multiple CO isotopologues (particular $^{12}$CO) would help with understanding the abundance and spatial distribution of gaseous material in the cavity, and resolving this peculiarity.}

\section{Discussion}

\subsection{Spiral Arms} \label{sec:disc_spirals}
The spiral arm morphology increases in complexity when eccentricity is introduced.
We see from Figure \ref{fig:sd_plot} and the explanation in Section \ref{sec:gsd} that distinguishing between the primary and secondary outer wakes becomes challenging when the companion is on an eccentric orbit. The radius and amplitude of the secondary wake is enhanced compared with the $e=0$ case. We expect that, just as in the $e=0$ case, the amplitude of both wakes will increase with mass. 

It is worth remarking on the rate of the rotation of the spiral arms, and whether they are in tension with the measured rotation rate of $0.6^{+3.3}_{-0.6}$ arcseconds per year by \cite{ren2018}. The one-sigma uncertainties indicate the rate of rotation is consistent with no rotation at all on one limit, or one rotation every 110 years at the other limit. The orbital period of our companion is approximately 160 years, and is therefore consistent with the constraints from \cite{ren2018}. 

On closer inspection of Figure \ref{fig:sd_plot} we can see that the apparent rotation rate of the spirals can be much less than the orbital rate of the companion, between $270^{\circ} \leq \nu \leq 90^{\circ}$ for example. Instead of co-rotating with the companion, the spirals appear to orbit once for every two orbits the companion. Rather than constraining the rotation rate of the spiral arms, observing changes in the spiral structure may be a better test of our scenario.



\subsection{Illumination Effects}\label{sec:ilum}

The greatest source of disagreement between our model and the observations of MWC 758 is the illumination of the outer disc regions. However we note that the polarised-intensity images presented in \cite{grady2013} do show extended emission behind the spiral arms. In particular, their H-band ($\lambda = 1.635\ \mu$m) PDI images show extended emission primarily behind S1, which is also the case in our model. It is possible that the Subaru/HiCIAO $H$-band observations in \cite{grady2013} are more sensitive to faint disc signal at large separation than the VLT/SPHERE observations. The Subaru observations were obtained at significantly lower airmass (1.1 vs. $>$1.9) and benefited from better adaptive optics correction at longer wavelength (1.6~$\mu$m instead of 1.04~$\mu$m). Furthermore, the smaller optical depth at longer wavelength may also contribute to the identification of features further from the star owing to less disc self-shadowing. This also seems to be the case for the $L'$ images which show more spiral structures at larger separation. 

The visual appearance of the spiral structure in our simulations in scattered light is very sensitive to illumination effects. To demonstrate this, we discuss how the illumination changes while increasing the dust opacity, \newc{by increasing the dust mass}, in Section \ref{sec:scattering}. We find that increasing the dust opacity results in an increased amount of scattering inside the gas-depleted cavity (see Figure \ref{fig:gtd}). Additional changes to the illumination pattern of the disc will also occur depending on the scale-height variations introduced by the spiral arms.



All of the scattered light features are produced in the surface layer of the disc, where heating effects from the central star are important. Since our simulation assumes the disc temperature only depends on radius, and not the height above the disc, the surface layers of our disc will be cooler than what they would be in reality. The reduction in temperature means that the surface layers are denser than they would be if heated by the primary star. \newc{This could affect the pitch angle of the spiral arms in our simulation. \cite{juhasz2018} showed that the pitch angle of spiral arms in the warm surface layers is higher than in the cold midplane. A higher pitch angle due to the faster propagation of spiral arms in a warm surface will affect the pitch angle of the spirals in our simulation. This could mean that we require a companion with a lower eccentricity than in our simulation, since our results indicate that an eccentric companion leads to higher pitch angles in the spiral arms close to the companion.}

Spiral features closer to the central star will intercept and block radiation from reaching more distant regions of the disc. The temperature inside the spiral shocks will also be higher than the surrounding material. This will change the appearance of the two bright spiral arms in our simulation, but how they will change is not known. It may have the effect of increasing the prominence of the two brightest spirals in our simulation, while dimming features outside of them, but this should be verified with radiative hydrodynamical simulations. \cite{lee2015} find that spiral arms in a vertically stratified disc have a lower contrast ratio in scattered light images than in an isothermal disc, but do not take shadowing effects into account. Vertical settling of micron-sized dust grains in the outer region could also dim the appearance of this region in scattered light \citep{Dullemond2004}.

The sharp decrease in polarised light outside of the spiral arms seen in the Y-band PDI image by \cite{benisty2015} is not seen in the H-band PDI image from \cite{grady2013}. The reason for this is unclear. The highly isotropic nature of the PDI images in both cases indicate that the scattering surface at these wavelengths is dominated by small dust grains. The precise dependence on polarisation with wavelength is entirely determined by the properties of these dust grains, and the aggregates that make them up \citep{volten2007}. Thus our inability to perfectly recreate the polarised intensity images from \cite{benisty2015} may be a result of our simplified dust grain model. 

To demonstrate that the main discrepancy of our model compared to the observations are simply a result of illumination effects, Figure~\ref{fig:dens_vs_scat} shows a comparison between the surface density of our simulation and the resulting Y-band PDI image where we artificially shadow the outer regions of the disc. This is to mimic the effects we believe are occurring in the Y-band PDI image from \cite{benisty2015}, and suggested by \cite{grady2013}. Comparing top right and bottom left panels in Figure~\ref{fig:dens_vs_scat}, the match is indeed much closer. Further, the K' band intensity images from \cite{grady2013} also show spiral-like structures outside of S1 and S2. The additional spiral arm in the south of the image, labelled S3, is seen in our Y-band PDI models, but not seen in the Y-band PDI observations, which is likely due to S1 shadowing S3 in the Y-band. \newc{Additionally, the inner wakes of our proposed companion may be responsible for the apparent shadowing on the tip of S2 see in the Y-band PDI observations.}

\subsection{Inclined or Eccentric Inner Disc?}

Our simulation predicts deviations from Keplerian rotation of the velocity profile inside the cavity, however we do not reproduce the sharp inner twist in the CO emission. There are two scenarios where our model may be able to produce this. Firstly, if our companion were initiated with some inclination, the inner disc is expected to also become inclined \citep{zhu2019}. Secondly, if the inner disc becomes more eccentric, the gas flow can appear to become more radial, which may also produce the twist, as suggested by \cite{boehler2018}. Better characterising velocity profile of gas inside the cavity of MWC 758 will help with refining our model, since an eccentric or inclined inner disc imply different properties for a potential companion. For example, an inclined companion can incline the inner disc and may produce the twist we see \citep{zhu2019}. 

It was recently suggested by \cite{francis2020} that the 3 au dust disc in MWC 758 is inclined roughly 30$^\circ$ with respect to the outer disc. We might then expect to see \newc{two prominent shadows} on the outer disc (e.g. across the spiral arms), as is occurring in other systems with an inclined inner disc, such as HD~142527 \citep{marino2015, francis2020} and HD~100453 \citep{benisty2017}. \newc{However there does not appear to be such obvious shadowing in MWC~758.} \newc{As mentioned in Section \ref{sec:com_nir}, there may be some shadowing of the inner disc onto the outer spirals, which could lead to the creation of the bright spot (feature v) in the top left panel of Figure \ref{fig:qphi}. There are two possible explanations for this structure. Firstly, it could be caused by a slightly inclined inner disc, such that feature v is the closer side of the inner disc with respect to our line of sight \citep{nealon2019}. Secondly, feature v could be a local enhancement in the gas surface density, as in our model, where a higher gas surface density leads to a higher elevation of the scattering surface, and hence shadowing of the outer disc.}

\newc{At first consideration, an inclined inner disc leading to feature v may seem consistent with scenario by \cite{francis2020}, however in their scenario the position angle of the disc is nearly pointing north (PA $=7 \pm 9^{\circ}$). Therefore if feature v is actually from this inclined inner disc, we should see the bright spot offset $90^{\circ}$ from the PA, so towards the East. The more subtle shadowing (between PA $\sim 180^{\circ}$ to $315^{\circ}$ in the bottom left panel of Figure~\ref{fig:rphi}) we discuss in Section~\ref{sec:com_nir} on the outer spirals is likely arising due to additional substructures closer towards the central star, and not from an inclined inner disc.}

In our simulations the inner gas disc is eccentric, which we have shown is capable of producing \newc{similar perturbations in moment~1} to what is seen in MWC~758 \citep{boehler2018}. It may then be possible that the inner mm dust disc is also eccentric, although we will need to verify this with dusty hydrodynamical simulations. If the dust disc is eccentric, it could be mistaken for an inclined inner disc. \newc{We also expect our eccentric inner disc to have its semi-major axis pointing towards the North, consistent with the measurements from \cite{francis2020}. Along with this, we can simultaneously explain the bright spot (feature v) in the $Y$-band PDI image. This feature is arising at the apocentre of the eccentric inner disc.}

\subsection{Companion Candidates}

The companion candidate found by \cite{reggiani2018} is located at $r = 0\farcs112$ from the central star with a position angle PA$\sim 170^\circ$. 
This is in slight tension with the companion in our model, which is $0\farcs2$ from the primary with a position angle of $210^{\circ}$. Our companion is significantly further from the primary star when we best match the spiral structure in MWC 758. Exploring the companion parameter space may yield a better match to the companion location, while still maintaining the match with the spiral structure. 

Another explanation for the point source is that it traces a local enhancement in the inner part of spiral S2, where it merges with the inner part of S1a. It is known that aggressive ADI processing of observations can cause extended emission, such as spiral arms, to appear as point sources \citep[e.g.][]{Christiaens2019}. Comparison between the bottom panels of Figure \ref{fig:qphi} shows indeed a spiral feature in the simulation at the location of `b' in the observation. If the spiral signal showed some azimuthal asymmetry at the location of `b' compared to the prediction from the simulation, the PCA-ADI algorithm would then convert it into a blob. Since the inner part of S2 appears to trace an accretion stream (see bottom left panel of Figure~\ref{fig:sd_plot}), such asymmetry could arise from a local enhancement in the accretion stream.

Our proposed scenario is in tension with the claimed detection of a planetary ($2-5$ M$_\textrm{J}$) companion connected with the south-eastern arm (S2) by \cite{wagner2019}. The separation of this proposed companion is almost co-located with the edge of the mm dust disc, and well within the $^{13}$CO and C$^{18}$O emission. A companion massive enough to be driving both of the spiral arms in MWC 758 should also be massive enough to clear a gap in the gas, or truncate the outer disc \citep{dong2018}. Given that the $^{13}$CO and C$^{18}$O emission observations from \cite{boehler2018} do not show any evidence of a gap or disc truncation at this separation, we suggest that the detection from \cite{wagner2019} is actually an extension of S2. Our proposition is readily testable with deeper observations in the near-IR of MWC 758.

\subsection{Observational Limits on Close Companions}

Several recent works in the literature have attempted to find the putative spiral-arm driving companions in MWC~758. Using VLT/NACO sparse aperture masking observations, \citet{grady2013} concluded that MWC~758 does not contain any stellar or brown dwarf ($\sim$80~M$_{\textrm{J}}$) companions within 300~mas. Their actual limit is likely slightly lower considering their use of a larger distance than measured by Gaia \citep[200 pc or 280 pc instead of 160 pc;][]{gaia2018}. \citet{reggiani2018} detected a point-like source located at 111 mas South from the central star, in a location similar to the position of the companion in our simulation. 

Using SPHERE/ZIMPOL, \cite{huelamo2018} constrained the H$_\alpha$ luminosity of any accreting sources in the disc. We found that the accretion rate of our companion is strongly time-dependent, ranging from $\sim$1$\times 10^{-9}-3\times 10^{-8}$ M$_\odot$/yr during the orbit of the companion. The accretion rate shows two peaks during the companion orbit, one at periastron, and once just after apastron. At the time we match the spiral structure of MWC~758 the accretion rate is $\sim$2$\times 10^{-9}$ M$_\odot$/yr. These values should be taken as upper limits, however, since the accretion rate is a function of the size of the sink particle. Using this number and assuming the circumplanetary disk accretion models from \cite{zhu2015b} and a planet radius of 2 R$_\textrm{J}$, the accretion luminosity of this source would be $L_{\textrm{acc}}\sim 7\times 10^{-4}$ L$_\odot$. \newc{We note that the circumplanetary disc models from \cite{zhu2015b} are constructed for planets on circular orbits. The size, and hence luminosity, of the accretion disc around a companion on an eccentric orbit would likely change substantially as the size of the planetary Hill sphere changes during the orbit.}

This luminosity is a factor 2 higher than the constraint presented in \citet{huelamo2018}. Mitigating this is that our radiative transfer calculations indicate that the line of sight optical depth of the companion can be above unity. The accretion luminosity constraints from \citet{huelamo2018} assume that the extinction of the companion will be the same as the central star, with an extinction of $A = 0.12$ mag. The optical depth when we best match the spiral structure is quite low however, at roughly 0.1, which corresponds to an extinction of $\sim$ 0.2 mag when adding the extinction towards the line of sight of the star (which is not included in our optical depth estimate). 

The optical depth is sensitive to the assumed properties of the dust grains, and a lower gas-to-dust ratio significantly increases the optical depth. Therefore before we can accurately quantify the line-of-sight extinction to the companion, we must understand the distribution and microscopic properties of the dust grains responsible for the near-IR scattering. Comparing our match to the Y-band PDI image of MWC~758 in Figure \ref{fig:qphi}, we can see that there is not as much scattering occurring inside the cavity of our simulation than in MWC~758. Increasing the scattering in the cavity would also increase the optical depth to the companion.

\subsection{Dust Morphology at mm Wavelengths}
\newc{Our scenario does not require the presence of an external perturber to reproduce the observed spiral morphology, however it is not yet clear whether it can reproduce the peculiar mm dust morphology in MWC~758 \citep{dong2018, boehler2018, casassus2019}.
Whether the outer dust asymmetry can be obtained without an external companion is a possibility that remains to be validated with dedicated simulations.}


Given the low mass of our companion, we expect that the mm dust morphology will exhibit a ring-like structure, similar to what has been shown in previous works \citep{pinilla2012}, and contrary to what occurs for stellar binaries \citep{ragusa2017,price2018,calcino2019,poblete2019}. However how the dust distribution changes with the introduction of eccentricity to planetary mass companions is largely unexplored. In general, the dust coupling to spiral structure, and thus the visual appearance at mm wavelengths, depends on the Stokes number \citep{veronesi2019}. \newc{Since eccentric planetary companions are also accompanied with eccentric gas discs \citep{kley2006}, it is expected that the dust cavity will also be eccentric, as seen in MWC 758 \citep{boehler2018, dong2018}}. A thorough treatment of dust grains, ranging across several orders of magnitude in Stokes numbers, is required to validate our model against observations at mm wavelengths.

\section{Conclusions}
In this paper we have examined the hypothesis that the transition disk MWC~758 hosts an unseen planetary mass companion on an eccentric orbit inside the CO cavity. We found that this scenario can explain several observational features in MWC758, such as
\begin{enumerate}
    \item the double spiral arm morphology,
    \item additional substructures seen in scattered light,
    \item the cavity in CO isotopologes, 
    \item the twist in the inner region of the velocity field.
\end{enumerate}

Our scenario makes several testable predictions which should be detectable with increased sensitivity in observations. Most notably we expect 
\begin{enumerate}
    \item additional spiral structure will be seen with improved sensitivity,
    \item the eastern and western spirals are not attached to an outer companion,
    \item a point source should be seen approximately 0.2" west/south-west of the central source,
    \item proper motion for the spiral arms consistent with a binary orbital period of $\sim 2\times 10^2$ yr should be detectable.
\end{enumerate}



\section*{Acknowledgements}
We thank the anonymous referee for their useful suggestions that improved the quality of this manuscript. JC acknowledges support from the Australian Government Research Training Program.
DP, CP and VC are grateful for Australian Research Council funding via FT130100034, FT170100040 and DP180104235. This project has received funding from the European Union's Horizon 2020 research and innovation programme under the Marie Sk\l{}odowska-Curie grant agreement No 210021. We acknowledge use of the OzStar supercomputing facility, funded by Swinburne University and the Australian Government.

\section*{Data Availability Statement}
The {\sc Phantom} SPH code is available from \url{https://github.com/danieljprice/phantom}. {\sc MCFOST} is available for use on a collaborative basis from \url{https://ipag.osug.fr/~pintec/mcfost/docs/html/overview.html}. The input files for generating our SPH simulations and radiative transfer models are available on request.




\bibliographystyle{mnras}
\bibliography{mwc758}



\section{Appendix}

\begin{figure*}
    \centering
    \includegraphics[width=1\linewidth]{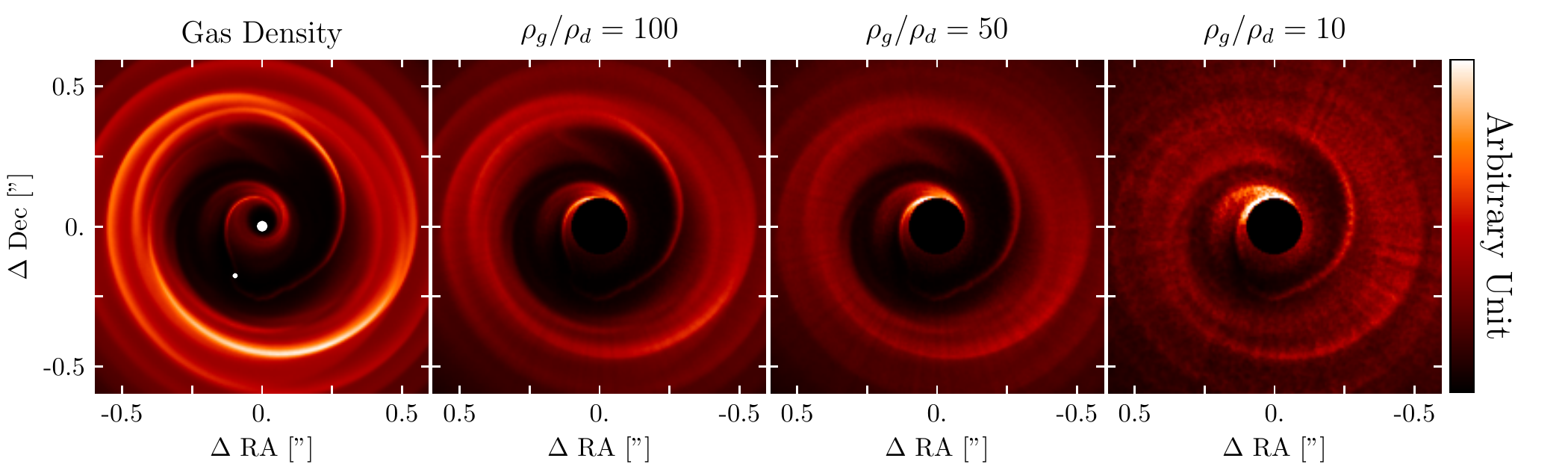}
        \caption{The gas surface density and the $1\,\mu$m scattered light intensity images of one timestep of our simulation for different values for the gas to dust ratio $\rho_\text{g}/\rho_\text{d}$. The images have been $r^2$ scaled to compensate for the $r^2$ dependence on the illumination of the disc surface. The appearance of the disc changes markedly when increasing the amount of dust (and hence opacity) in the disk. The spirals inside the cavity become more pronounced, while the presence of the cavity becomes less apparent. The true anomaly of the companion is $\nu = 107^\circ$ at this snapshot. The colorbar on the $\rho_\text{g}/\rho_\text{d}=50$ and $\rho_\text{g}/\rho_\text{d}=10$ panels are scaled by a factor of 1.25 and 2 compared to the $\rho_\text{g}/\rho_\text{d}=100$ panel, respectively.}
    \label{fig:gtd}
\end{figure*}

\subsection{Scattering Dependence on the Dust Opacity} \label{sec:scattering}

Figure \ref{fig:gtd} shows the $r^2$-scaled scattered light intensity of a single timestep of our simulation for different values of the gas to dust ratio. The appearance of the spiral arms inside the cavity is sensitive to the opacity of the disc at short wavelengths, which is almost entirely determined by the small dust grain population. As we increase the amount of dust (and hence opacity), the central cavity changes from being optically thin to optically thick and begins scattering the photons from the central source. 

As the cavity becomes optically thick to infrared radiation the spiral structure becomes more apparent and the prominence of the cavity diminishes. The spirals also begin to shadow more distant regions of the disc. A similar process is likely occurring in MWC~758. Some structure observed in this disc is within the cavity seen in CO isotopologes \citet{boehler2018}, which roughly probe the gas surface density. A reduction in CO in the cavity implies there is a reduced amount of gas inside the cavity. However structure is still seen inside this CO cavity in the scattered light observations of \citep{benisty2015}. This indicates that despite the reduction in gas, the inner cavity still maintains a high opacity at near-infrared wavelengths. 

For the cavity in our simulation to remain optically thick requires a low gas to dust ratio, on the order of $\sim$10:1. As the dust fraction is increased, the scattering surface of the infrared photons also increases across the disc. This presents a challenge for SPH simulations, since the best resolved portions of a simulation are those with the highest density (i.e. the mid-plane). The low SPH resolution becomes apparent when $\rho_\text{g}/\rho_\text{d} = 10$, where individual Voronoi cells can be distinguished 


Low SPH resolution close to the central sink particle also presents challenges for our ray-tracing. In this region, individual Voronoi cells close to the central sink can become optically thick, casting a shadow over the outer region of the disc. This effect, and the previous, are expected to disappear if the resolution of the SPH simulation was increased further. 

The gas disc inside the orbit of the companion, and around the primary star, accounts for an increased amount of the total disc luminosity as the gas to dust ratio is decreased. This presents a challenge when comparing our radiative transfer models to MWC 758. Although there is abundant scattering close to the coronagraph in the scattered light images presented by \cite{benisty2015}, the scattering in our model is much higher. We expect that this is occurring because there is a gas cavity close to the primary star (see left-most panel of Figure \ref{fig:gtd}). The photons from the central star propogate in near vacuum until encountering the optically thick wall of the inner cavity. This inner cavity arises due to the properties of our central sink particle, and the resolution of the simulation. Decreasing the size of the sink particle would mitigate this effect, but would come with significant computational cost. To mitigate this effect we apply a Gaussian taper to the inner disc region to reduce its luminosity when we conduct our post-processing of the infrared images.

\bsp	
\label{lastpage}
\end{document}